\documentclass[prc,twocolumn,superscriptaddress,showpacs,preprintnumbers,amsmath,amssymb,nofootinbib]{revtex4-1}
\usepackage[dvips]{graphicx}
\usepackage{color}
\usepackage{ulem}
\def\m@thcombine#1#2{%
  \setbox0=\hbox{$#1$}
  \setbox1=\hbox{$#2$}
  \ifdim\wd0>\wd1
    \setbox0=\hbox to\wd1{\hss\box0\hss}
  \else
    \setbox1=\hbox to\wd0{\hss\box1\hss}
  \fi
  \mathop{\vcenter{
    \offinterlineskip\box0\box1}}}
\def\lesim{\m@thcombine<\sim}
\def\gesim{\m@thcombine>\sim}

\def\vr{\mbox{\boldmath$r$}}

\def\+{\mbox{\unboldmath $+$}}
\def\-{\mbox{\unboldmath $-$}}
\def\={\mbox{\unboldmath $=$}}

\begin{document}

\title{Coexistence of Anderson-Bogoliubov phonon and quadrupole cluster vibration in neutron star inner crust 
}

\author{Tsunenori Inakura}
\affiliation{Department of Physics, Faculty of Science, Niigata University, Niigata 950-2181, Japan}
\affiliation{Laboratory for Advanced Nuclear Energy, Institute of Innovative Research, Tokyo Institute of Technology, Tokyo 152-8550, Japan}
\author{Masayuki Matsuo}
\affiliation{Department of Physics, Faculty of Science, Niigata University, Niigata 950-2181, Japan}

\begin{abstract}
\noindent {\bf Background:}
Low-lying collective excitations of inner crust matter in neutron stars are expected to have impact on observables
such as the quasi-periodic oscillation in giant flares and the cooling of inner crust in transient phenomena. Coupling
between Anderson-Bogoliubov superfluid phonon (AB phonon) in superfluid neutron gas and collective excitations of
nuclear clusters is a key issue.
\\
{\bf Purpose:}
We intend to predict the nature of low-lying excitation modes of superfluid inner crust matter with focus on
quadrupole excitations around a spherical nuclear cluster.   We study how the AB phonon of neutron superfluid
and possible quadrupole shape vibration of clusters are affected by the inhomogeneous structure of inner crust matter.
\\ 
{\bf Methods:} The coordinate-space Hartree-Fock-Bogoliubov method and the  quasiparticle random-phase approximation formulated in a spherical Wigner-Seitz cell are adopted to describe neutron 
superfluidity and low-lying collective excitations. We perform systematic numerical calculations
for quadrupole excitations by varying the neutron chemical potential and the number of protons in a cell. \\
{\bf Results:} 
The calculated results indicate emergence of both AB phonon and quadrupole shape vibration of the cluster
with small mixing between the two collective modes.
The quadruple AB phonon is similar to that in uniform superfluid apart from small admixture of the shape vibration of cluster. 
The excitation energy and the collectivity of the cluster vibration mode shows strong and oscillatory dependence on the
neutron chemical potential (the neutron gas density), leading to softening and instability in certain situations.
This is caused by the resonance shell effect where unbound but resonant 
single-particle states of neutrons play a central role. \\
{\bf Conclusions:} 
The AB phonon of superfluid neutron gas and the quadrupole shape vibration of nuclear cluster coexist in inner crust.
The coupling between the AB phonon and the quadrupole shape vibration is weak.
The collectivity of the quadrupole shape vibration is governed by the resonance shell effect, and it suggests that
emergence of deformed nuclear clusters is possible at any layer of inner crust.
\end{abstract}
\maketitle

\section{Introduction}

Nuclear matter in the inner crust of neutron stars consists of a lattice of nuclear clusters immersed in superfluid neutron gas\cite{Chamel-Haensel2008,Haensel-book,Pethick-Ravenhall95}.
The coexistence of nuclear clusters and superfluid neutrons brings about rich many-body phenomena. 
A typical example might be the pinning and unpinning of superfluid vortices, which is considered to be responsible for the glitch in
the rotation frequency of a neutron star\cite{Anderson-Itoh1975,Alpar1977,Pines-Alpar92}. Another interest is found in collective excitations of inner crust matter. 
The lattice vibration, the collective displacement motion of nuclear clusters, may be  microscopic
entity of the seismic oscillation observed as the quasi-periodic oscillation in the giant flares\cite{Duncan1998,Samuelson2007,Andersson09}. 
Recently the superfluid phonon\cite{Anderson58,Bogoliubov59,Galitskii58}, called also the Anderson-Bogoliubov 
phonon,
attracts new attentions\cite{Aguilera09,Pethick10,Cirigliano11,Page-Reddy2012,Kobyakov13,Kobyakov14,Chamel13,Martin14}.
It may influence the heat capacity (due to this additional degrees of freedom)\cite{Khan05,Martin14},  
the seismic oscillation  (through the coupling between the superfluid and the lattice phonons)\cite{Pethick10,Chamel13},
and the thermal conductivity (as a new heat carrier)\cite{Aguilera09,Pethick10,Cirigliano11,Page-Reddy2012}. 
Possibility of forming a exotic crystalline structure
is also suggested\cite{Kobyakov13,Kobyakov14}. 
The coupling between the nuclear cluster and the superfluid phonon is regarded as a key issue in these studies.

Theoretical approaches to the collective excitation of inner crust matter have been undertaken in
two different directions. One is macroscopic approaches focusing on low-frequency limit\cite{Aguilera09,Pethick10,Cirigliano11,Page-Reddy2012,Kobyakov13,Kobyakov14,Chamel13,Magierski04,Martin16}, 
where the collective degrees of freedom of the superfluid phonon is explicitly introduced
 in the thermodynamic limit or in the hydrodynamic framework 
while the nucleon degrees of freedom are treated implicitly.
Another is microscopic approaches based on the many-body quantum mechanics for interacting
nucleons.  In the latter direction
the linear response methods (the quasiparticle random phase approximation QRPA) based
on nuclear density functional models or the Hartree-Fock-Bogoliubov (HFB) models  are often adopted\cite{Khan05,Grasso08,Baroni2010,Martin14,Inakura17}.
Properties of the superfluid phonon mode are studied in detail within the QRPA 
for pure and uniform neutron superfluid\cite{Martin14}. However, if one intends to describe
 the inhomogeneous system consisting of 
nuclear cluster and neutron superfluid,  one resorts to QRPA calculations  
assuming a spherical Wigner-Seitz cell enclosing a single cluster\cite{Khan05,Grasso08,Baroni2010,Inakura17}. 
 In a preceding publication (Part I hereafter)\cite{Inakura17} 
we have studied along this line the dipole excitation
in the cell since  the displacement motion of cluster has the multipole $L=1$ and
the dipole channel is responsible for the coupling
between the superfluid phonon and the lattice phonon. This study predicts that
 the superfluid phonon with the dipole mulitpolarity 
is influenced strongly by the presence of cluster in such a way that the amplitude of superfluid phonon
is significantly hindered inside the nuclear cluster. This result suggests that the
coupling between the superfluid phonon and the displacement motion of cluster
(i.e. the lattice phonon) might be weak, implying possible enhancement of the
thermal conductivity of the inner crust.

In the present work we analyze the quadrupole excitation in the spherical 
Wigner-Seitz cell of inner crust with use of the same model as that in Part I. 
In a pioneering work by Khan et al., where a HFB plus QRPA model 
is applied to quadrupole excitations, a low-lying
collective mode called the supergiant resonance is predicted\cite{Khan05}. Its relation
to the superfluid phonon is suggested, but only in a qualitative argument.
In another study using a Hartree-Fock plus RPA, in which the nucleon pair correlation is
neglected and the superfluid phonon is not expected, 
low-lying collective modes with a character of the multipole shape vibration of cluster are predicted\cite{Baroni2010}.
Because of this situation, a consistent picture is not obtained yet
and questions and problems remain.    
What is the relation among
the superfluid phonon, the supergiant resonance in Khan et al.\cite{Khan05} and
the shape vibration of cluster? 
How is the superfluid phonon affected by the presence of cluster? Is there difference between
the dipole and quadrupole modes of the superfluid phonon?
Does the cluster shape vibration exist or how is it affected when it is immersed in  superfluid neutron gas? 
The purposes of the present study is to answer these
 questions.  
In short, we intend to  clarify the nature of the low-lying excitation modes of superfluid inner crust matter
with focus on quadrupole excitations around a spherical nuclear cluster.

In the following we call the superfluid phonon mode of neutron
superfluid  the Anderson-Bogoliubov phonon mode  (abbreviated as AB phonon) to keep the same nomenclature as
in Part I\cite{Inakura17}.
The paper is organized as follows: 
In Sec. II, we briefly explain the model we use to describe the ground state and the excitation modes of inner crust matter in the
Wigner-Seitz approximation. In Sec. III, 
we show numerical results of the QRPA calculation performed for the quadrupole excitations in the spherical Wigner-Seitz
cell. We will show that there emerge two low-lying collective excitations, which are related to
both the AB phonon mode of superfluid neutron gas
and the quadrupole shape vibration of the nuclear cluster.  We will investigate in detail basic properties of these
low-lying collective modes to clarify the interplay between the AB phonon mode and the shape vibration of cluster.
It will be turned out that the quadrupole shape collectivity of the cluster exhibits a peculiar shell effect which arises
not from neutrons bound to clusters, but from unbound neutrons permeating the nuclear cluster and neutron gas.
Sec. IV is devoted to the conclusion.

\section{Model}

We employ the HFB model and the QRPA theory. 
The calculation framework and the model parameters are 
the same as those in Part I\cite{Inakura17}.
Here we recapitulate the adopted HFB+QRPA formalism briefly.
For other details, we refer the readers to Refs.~\cite{Inakura17,Matsuo01,Serizawa09,Matsuo10,Nakatsukasa2016}.

To calculate the properties of the inner crust, we introduce the spherical Wigner-Seitz approximation
with a Wigner-Seitz radius of $R_\mathrm{WS}=20$~fm.
Following the standard prescriptions adopted in the HF/HFB 
calculations\cite{Negele73,Barranco1998,Sandulescu04a,Sandulescu04b,Baldo2005,Baldo2006,Grill11,Pastore11,Pastore12},
we impose the von-Neumann-Dirichlet boundary condition\cite{Negele73}.
Nucleon configuration of the Wigner-Seitz cell is specified with the neutron chemical potential (Fermi energy) 
$\lambda_n$  and a fixed integer number $Z$ of protons in the cell.

The ground state configuration
 of inner crust matter for given $Z$ and $\lambda_n$ is obtained 
by solving the HFB equation in the coordinate representation. 
We adopt the zero-range effective Skyrme SLy4 interaction~\cite{SLy4} for the Hartree-Fock 
potential (in the particle-hole channel).
A version of density-dependent delta interaction~\cite{Matsuo06,Matsuo07} is used to obtain the
 pairing potential (in the pairing channel)
\begin{eqnarray}
\Delta_\tau(\vr) = V_\mathrm{pair} \left[ 1 - \eta \left(\frac{\rho_\tau(\vr)}{\rho_c}\right)^\alpha \right] \tilde{\rho}_\tau(\vr)
\label{pairing.Delta}
\end{eqnarray}
and the pair density
\begin{eqnarray}
\tilde{\rho}_\tau(\vr)=\langle \Psi_0 | \psi     (\vr\uparrow)   \psi     (\vr\downarrow) | \Psi_0 \rangle.
\end{eqnarray}
Here $\tau=n,\, p$ denotes isospin, $\uparrow$ and $\downarrow$ are up and down components of spin $\sigma$, and $ \rho_c=0.08$ fm$^{-3}$.
We set the cut-off $j_\mathrm{max}= (75/2)\hbar$ for the single-particle partial waves and we include all the quasiparticle states under the cut-off quasiparticle energy $E_\mathrm{cut}=$ 60 MeV. To characterize the neutron superfluid
outside the nuclear cluster
we evaluate its average neutron density and average pair potential
 as
\begin{eqnarray}
\rho_\mathrm{ext}   = \frac{\int^{R_2}_{R_1} dr \, r^2 \rho_n(r)  }{\int^{R_2}_{R_1} dr \, r^2} \,,\quad
\Delta_\mathrm{ext} = \frac{\int^{R_2}_{R_1} dr \, r^2 \Delta_n(r)}{\int^{R_2}_{R_1} dr \, r^2}
\label{matter.Delta.rho}
\end{eqnarray}
with $R_1 =$ 12 fm and $R_2 =$ 18 fm.

The QRPA calculation is performed on top of the HFB ground state $\Psi_0$.
In the present study we focus on the quadrupole excitations.
We utilize the linear response formalism~\cite{Matsuo01,Serizawa09,Matsuo10} for the QRPA. 
We describe responses of the system with respect to the quadrupole operator $Q$,  
and the pair addition and removal operators, $P_\mathrm{add}$ and $P_\mathrm{rm}$, defined by 
\begin{eqnarray} 
&& Q              = \sum_{\sigma} \int \!d\vr\, r^2Y_{2M}(\hat{\vr})\, \psi^\dag(\vr\sigma)\psi(\vr\sigma)  \,, \nonumber \\
&& P_\mathrm{add} =         \int \!d\vr\,  Y_{2M}(\hat{\vr})\, \psi^\dag(\vr\downarrow)\psi^\dag(\vr\uparrow) \,, \nonumber \\
&& P_\mathrm{rm}  =         \int \!d\vr\,  Y_{2M}(\hat{\vr})\, \psi     (\vr\uparrow)  \psi     (\vr\downarrow) \,.
\label{eq:operators}
\end{eqnarray}
We solve the QRPA linear response equations for fluctuations of the nucleon density $\rho(\vr)$, the 
nucleon pair density $\tilde{\rho}(\vr)$  and its complex conjugate $\tilde{\rho}^*(\vr)$. 
The spectral representation is adopted for the density response function and 
all the quasiparticle states adopted in the HFB calculation are included. 
We calculate the strength function 
\begin{eqnarray}
S(O;E) = \sum_{Mi} \delta(E - E_i) \left|\langle \Psi_i^{2M}| \hat{O} | \Psi_0\rangle\right|^2 \,,
\end{eqnarray}
for the quadrupole excited states $ \Psi_i^{2M} $ with respect to
the operators $\hat{O}=Q, P_\mathrm{add}$ and $P_\mathrm{rm}$
of neutrons and protons separately.
With a small imaginary
constant $\epsilon$ in the energy argument, the 
delta function peaks in 
the strength functions are smeared with the Lorentzian function having
 the FWHM of $2\epsilon$. 
We evaluate the strength 
$B(O)=\sum_M\left|\langle \Psi_i^{2M}| \hat{O} | \Psi_0\rangle\right|^2$
of each excited state by integrating the strength function in an energy interval $E \in [E_i - 10 \epsilon, E_i + 10 \epsilon]$ around its peak energy $E_i$. We employ $\epsilon=10$ keV.
Three transition densities from the HFB ground state $\Psi_0$ to the $i$th QRPA excited state $\Psi_i^{2M}$,
\begin{eqnarray}
&& \delta       \rho _\mathrm{ph}(\vr) =   \langle \Psi_0 | \sum_\sigma\psi^\dag(\vr\sigma)     \psi     (\vr\sigma)     | \Psi_i^{2M} \rangle = Y_{2M}(\hat{\vr})\delta \rho _\mathrm{ph}(r)  \,,\nonumber \\
&& \delta\tilde{\rho}_\mathrm{pp}(\vr) = \langle \Psi_0 | \psi     (\vr\uparrow)   \psi     (\vr\downarrow) | 
\Psi_i^{2M}  \rangle = Y_{2M}(\hat{\vr})\delta\tilde{\rho}_\mathrm{pp}(r) \,,\nonumber \\
&& \delta\tilde{\rho}_\mathrm{hh}(\vr) =  \langle \Psi_0 | \psi^\dag(\vr\downarrow) \psi^\dag(\vr\uparrow)   | 
\Psi_i^{2M}  \rangle  = Y_{2M}(\hat{\vr})\delta\tilde{\rho}_\mathrm{hh}(r)\,,
\label{eq:densities}
\end{eqnarray}
are obtained from the corresponding fluctuating densities at the peak energy $E_i$.
Note that the calculated spectrum is discretized because of the boundary condition.

As the residual interaction which enters in the QRPA calculation, 
we adopt the same effective pairing interaction for the pairing channel. 
Concerning the residual interaction in the particle-hole channel, 
we adopt the Landau-Migdal approximation~\cite{Serizawa09,Matsuo10,Khan04,Khan09,Paar07} with a renormalization
scheme often employed in this approximation. Namely we replace the
self-consistent particle-hole interaction $v_{\mathrm{ph}}$ by the 
Landau-Migdal interaction $f \times v_{\mathrm{LM}}$ derived from the Skyrme
interaction and renormalized with a factor $f$.
The factor $f$ is determined to bring a peak corresponding to the displacement motion
to zero energy in the drip-line nucleus for a given $Z$. 
The same value of $f$ is used for the Wigner-Seitz cells with the same $Z$.

\section{Results and discussion}

We have performed the HFB and QRPA calculations systematically for various configurations
specified by the neutron chemical potential $\lambda_n$ and
the proton number $Z$, with changing $Z$ and $\lambda_n$, in order to study basic properties of excitation modes of
 inner crust matter. 
The adopted proton numbers are $Z=$ 20, 28, 40, and 50, chosen to cover the range predicted
in the previous HFB or HF calculations for the equilibrium \cite{Negele73,Baldo2005,Baldo2006,Grill11}.
For the neutron chemical potential, we vary it in the range 
$\lambda_n = 1-6$ MeV, which corresponds to the density of neutron matter 
$\rho_n \approx 5 \times 10^{-4} - 1 \times 10^{-2}$ fm$^{-3}$. 
The ground states obtained by the HFB calculation are same as in Part I~\cite{Inakura17}.

\subsection{Low-lying quadrupole excitations}

\begin{figure}[tb]
\begin{center}
\includegraphics[width=0.490\textwidth,keepaspectratio]{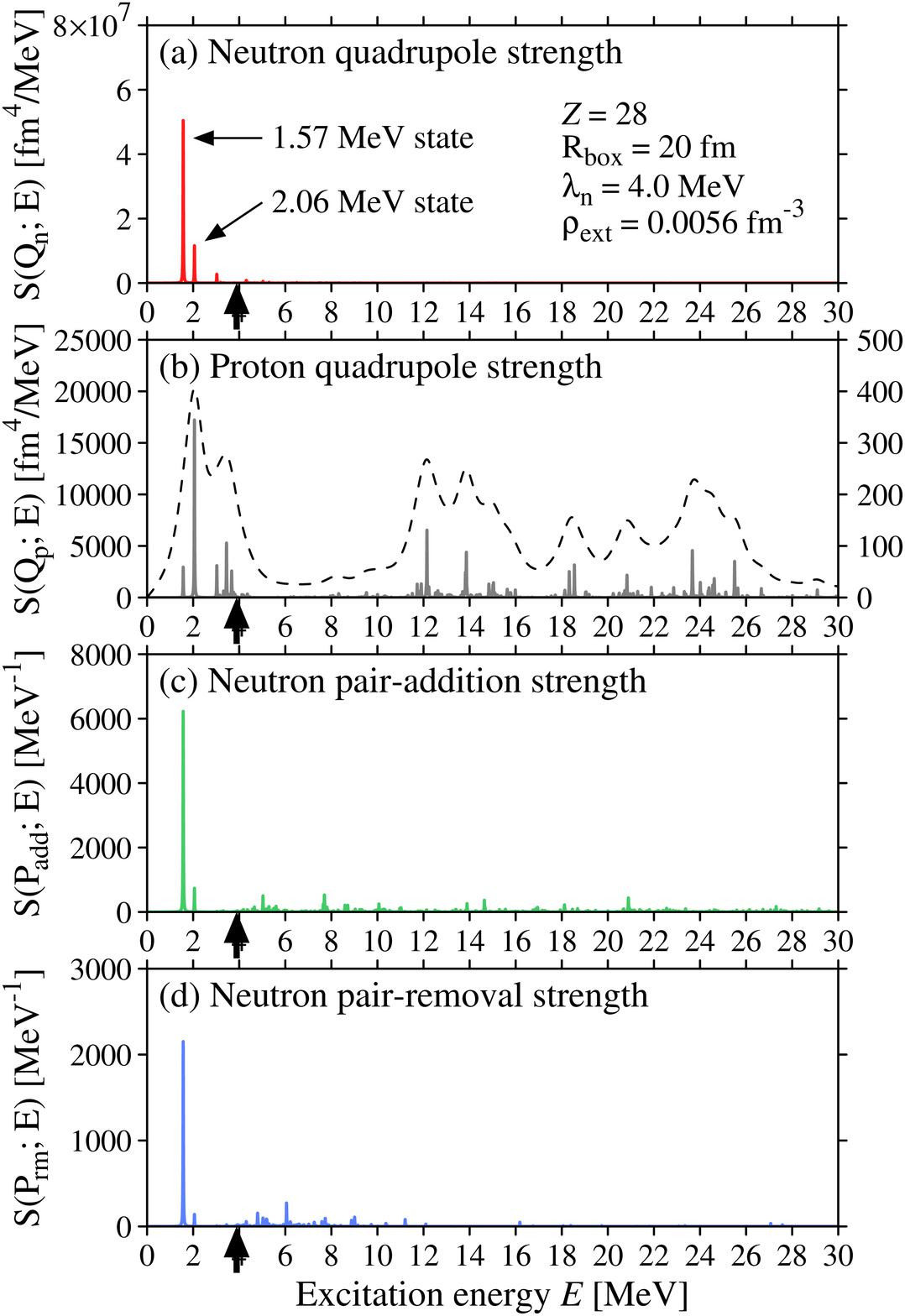}
\caption{(a) Strength function $S(Q_n;E)$ for neutron quadrupole operator, 
(b) $S(Q_p;E)$ for proton quadrupole operator, (c) $S(P_\mathrm{add};E)$ for neutron pair-addition operator  
and (d) $S(P_\mathrm{rm};E)$ for neutron pair-removal operator, calculated 
for $Z=28$ system with $\lambda_n =$ 4.0 MeV,
plotted as a function of the excitation energy $E$.
Arrows indicate the threshold energy $2\Delta_\mathrm{ext}$. 
The smearing parameter is $\epsilon=$ 10 keV. Dashed curve
in (b) is the proton quadrupole strength function  $S(Q_p;E)$ obtained with
$\epsilon=$ 500 keV (with the scale on the right axis).
}
\label{neutron.proton.Padd.Prm}
\end{center}
\end{figure}

We first discuss the case of $Z=28$ and $\lambda_n =$ 4.0 MeV  chosen as representative.

Figures~\ref{neutron.proton.Padd.Prm}(a)-(d) show the strength functions 
$S(O;E)$ for the neutron quadrupole moment, $O=Q_n$, the proton quadrupole moment  $Q_p$, 
the neutron pair-addition operator $P_\mathrm{add}$ and the neutron pair-removal operator $P_\mathrm{rm}$,
respectively, 
for the $Z=28$ system with $\lambda_n = 4.0$ MeV.  Two distinct peaks at $E=$ 1.57 MeV and
at $E=$ 2.06 MeV are seen in the low-lying region. 
These peaks are located below $2\Delta_\mathrm{ext}$ (twice of the neutron pair gap, indicated with arrows). 
Above this threshold energy, the QRPA excited states
are densely distributed (and would become continuous if $R_\mathrm{WS} \rightarrow \infty$). 
It is noted here that the peak at $E=$ 1.57 MeV
is significant in the strength functions 
for the neutron operators $Q_n$, $P_\mathrm{add}$  and $P_\mathrm{rm}$, but not in the
strength function for the proton operator $Q_p$. The second peak at $E=$ 2.06 MeV, however, 
has the largest strength in the proton quadrupole moment $Q_p$, but the strengths for
the neutron operators are much smaller than those of the first peak at $E=$ 1.57 MeV.
Apparently there exist two kinds of low-lying collective 
modes which have
different characters.

Let us examine the peak at $E=$ 1.57 MeV in detail.
This state has the strengths $B(Q_n)= 1.49 \times 10^6$ fm$^4$ , 
$B(P_\mathrm{add})= 183$ and $B(P_\mathrm{rm})= 63$ for the neutron operators, which 
are the largest among the QRPA excited states and exhaust
67 \%, 21 \%, and 42 \%, respectively,
of the total strengths (non-energy-weighted sums integrated up to 50 MeV).
Note that the neutron quadrupole strength $B(Q_n)=1.49 \times 10^6$ fm$^4$ is larger by a factor of  $10^{2-3}$
than a typical value of the low-lying collective $2_1^+$ state in isolated nuclei, 
$B(E2)/e^2 \sim 10 B_\mathrm{W}$ ($B_\mathrm{W}$ being the Weisskopf unit\cite{Ring-Schuck}).
Having this extraordinary quadrupole strength, we consider that this state may correspond to what is called
the supergiant quadrupole resonance in Khan et al.\cite{Khan05}. 

\begin{figure}[tb]
\begin{center}
\includegraphics[width=0.4850\textwidth,keepaspectratio]{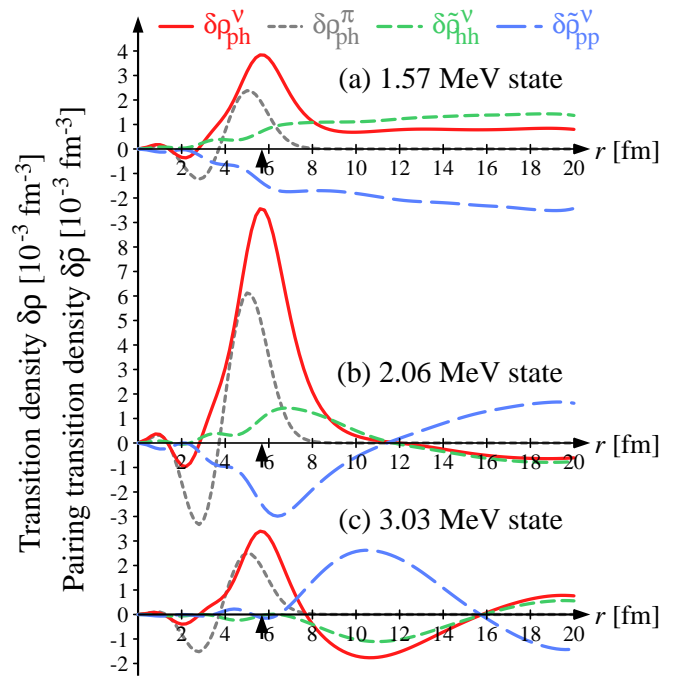}
\caption{Transition densities of the low-lying quadrupole collective modes whose excitation energies are
(a) $E=$ 1.57 MeV, (b) 2.06 MeV, and (c) 3.03 MeV, obtained for $Z=$ 28 system with $\lambda_n=$ 4.0 MeV. 
Red, gray, green, and blue curves are $\delta\rho^\nu_\mathrm{ph}(r)$, $\delta\rho^\pi_\mathrm{ph}(r)$, 
$\delta\tilde{\rho}^\nu_\mathrm{pp}(r)$, and $\delta\tilde{\rho}^\nu_\mathrm{hh}(r)$, respectively. 
The arrows indicate the surface position of the cluster\cite{Inakura17}. See text for details. 
}
\label{drho}
\end{center}
\end{figure}

Figure~\ref{drho}(a) shows the neutron and proton particle-hole transition densities,
$\delta\rho^\nu_\mathrm{ph}(r)$ and $\delta\rho^\pi_\mathrm{ph}(r)$, and the neutron pair-addition and -removal 
transition densities, $\delta\tilde{\rho}^\nu_\mathrm{pp}(r)$ and $\delta\tilde{\rho}^\nu_\mathrm{hh}(r)$,
of the 1.57 MeV state. It is seen that the neutron particle-hole transition density
$\delta\rho^\nu_\mathrm{ph}(r)$ as well as the neutron pair-addition/removal 
transition densities, $\delta\tilde{\rho}^\nu_\mathrm{pp}(r)$ and $\delta\tilde{\rho}^\nu_\mathrm{hh}(r)$,
have finite amplitudes in the whole region of neutron gas $r \gesim 6$ fm. This is the origin of the extremely large
neutron strengths. The three transition densities far outside the nucleus $r \gesim 9$ fm exhibits
a signature of the AB phonon mode: They share an approximate
sinusoidal shape in the neutron gas region while the pair transition densities $\delta\tilde{\rho}^\nu_\mathrm{pp}(r)$ and $\delta\tilde{\rho}^\nu_\mathrm{hh}(r)$ have an opposite phase to one another\cite{Inakura17}.
The neutron and proton particle-hole
transition densities  $\delta\rho^\nu_\mathrm{ph}(r)$ and $\delta\rho^\pi_\mathrm{ph}(r)$, however,
show enhanced amplitude in the surface region of the cluster, 4 fm  $\lesim r \lesim$ 8 fm, 
with the phase coherence. This behaviour in the surface region is similar to the one known in the
low-lying quadrupole shape oscillation of isolated nuclei. 

The 2.06 MeV state is recognized in the proton quadrupole
strength function, but it is hard to be seen in the neutron strength functions. 
The proton strength $B(Q_p)=$ 508 fm$^4$ of the 2.06 MeV state is the same order of magnitude as
the typical $B(E2)$ values of the low-lying quadrupole vibration in isolated nuclei. 
(Note $B(E2)=e^2B(Q_p)$.) The neutron quadrupole strength $B(Q_n)=3.46 \times 10^5$ fm$^4$
is two orders of magnitude larger than those in isolated nuclei, but significantly smaller than
that of the 1.57 MeV state. 
The transition densities of the 2.06 MeV state, shown in Fig.~\ref{drho}(b), displays a similar
behavior as those of the 1.57 MeV state, but some difference is clearly seen: The neutron
and proton particle-hole amplitudes around the
cluster surface is much larger than those of the 1.57 MeV state and it suggests a closer relation to the
quadrupole shape vibration of the cluster. The neutron amplitudes outside the cluster
displays a node at $r \approx$ 12fm.

\begin{figure}[tb]
\begin{center}
\includegraphics[width=0.4730\textwidth,keepaspectratio]{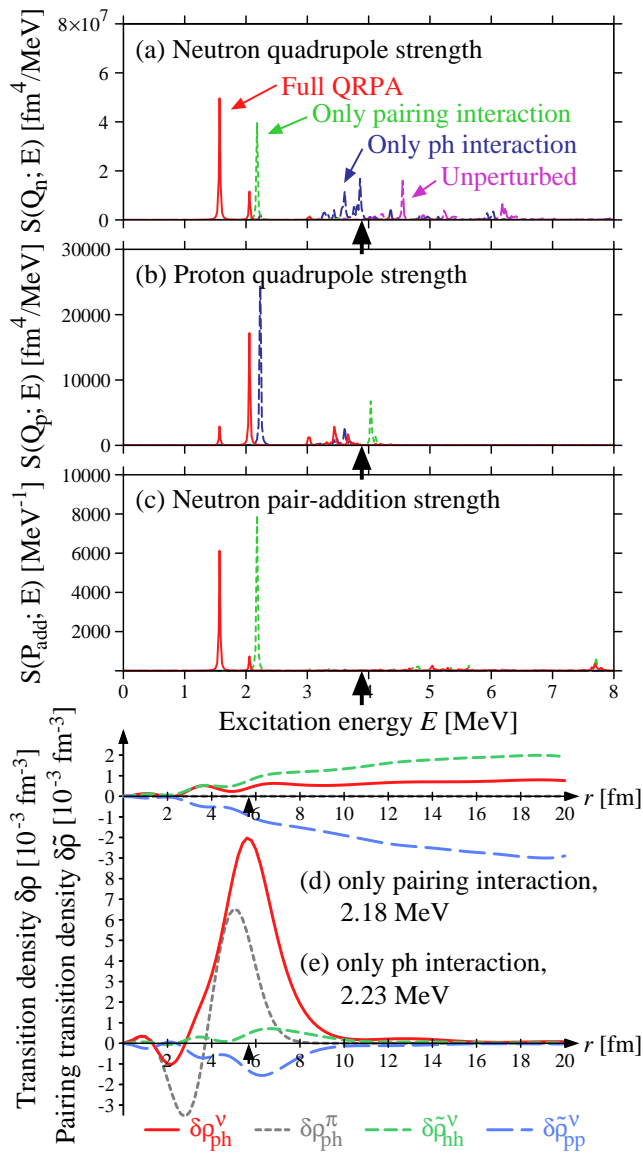}
\caption{
Uppers: (a) Strength function $S(Q_n;E)$, (b) $S(Q_p;E)$, and (c) $S(P_\mathrm{add};E)$ 
calculated with the full residual interaction (red curve), with only the residual pairing interaction (green dashed), and
with only the residual particle-hole interpretation (blue long-dashed).
Unperturbed results (magenta long-dashed dotted) are also plotted.
The system is  $Z=28$ and $\lambda_n = 4.0$ MeV.
Lowers: Transition densities of the low-lying collective mode obtained in calculations 
(d) with only residual pairing interaction and (e) with only the residual particle-hole interaction.
Red, gray, green, and blue curves are  $\delta\rho^\nu_\mathrm{ph(r)}$, $\delta\rho^\pi_\mathrm{ph}(r)$, 
$\delta\tilde{\rho}^\nu_\mathrm{pp}(r)$, and $\delta\tilde{\rho}^\nu_\mathrm{hh}(r)$, respectively. 
See also the captions of Figs.~\ref{neutron.proton.Padd.Prm} and \ref{drho}.
}
\label{residual}
\end{center}
\end{figure}

\subsection{Coexistence of AB phonon  and quadrupole cluster vibration}

In order to investigate the characteristic features of these two collective excitations,
we look into roles of the residual interaction.
For this purpose we shall drop off some parts of the residual interaction which enters in the QRPA calculation. 
Figure~\ref{residual} shows the strength functions $S(Q_n;E)$, $S(Q_p;E)$ and $S(P_\mathrm{add};E)$,
1) which are obtained using
only the residual pairing interaction (i.e. the particle-hole part of the residual interaction is neglected)
plotted with green curves, 2) those (blue curves) obtained with only the residual particle-hole interaction 
(the residual pairing interaction is neglected instead), and 3) unperturbed results (magenta curves) in which 
all the
residual interactions are neglected. 
The obtained strengths functions are shown in the panels (a), (b) and (c).
The calculation using only the residual pairing interaction produces 
a collective state which appears as a peak at 2.18 MeV
with large neutron strengths. This state has apparent characters of the
AB phonon mode as are seen
in its transition densities, shown in Fig.~\ref{residual}(d): 
the oscillatory 
behavior with long wave length, $\delta\tilde{\rho}^\nu_\mathrm{pp}(r) \sim  A j_2(qr) $ and 
$\delta\tilde{\rho}^\nu_\mathrm{hh}(r) \sim - A j_2(qr) $,
is clearly seen, especially outside the nuclear cluster $r \gesim 8$ fm.
When only the particle-hole interaction included (and the residual pairing interaction is neglected),
the quadrupole shape vibration of the cluster emerges seen as   
a peak at 2.23 MeV in the proton strength function. In fact, 
transition densities of this state [Fig.~\ref{residual}(e)] exhibit clearly features of
 isoscalar quadrupole shape vibration of the cluster:
The particle-hole transition densities $\delta\rho^\nu_\mathrm{ph}(r)$ and $\delta\rho^\pi_\mathrm{ph}(r)$ 
of neutrons and protons both have large amplitudes with the same phase
around the surface region of the cluster, 4 fm $\lesim r \lesim$ 8 fm, 
while all the transition densities diminish outside the cluster $r \gesim 10$ fm.

Having these observations, we can interpret the two excitation modes 
at $E=$1.57 MeV and 2.06 MeV in the full QRPA calculation 
 as mixtures or coupled modes of the AB phonon mode and the quadrupole shape vibration of the cluster. The transition
densities of the 1.57 MeV state, Fig.~\ref{drho}(a),  appear  an in-phase superposition of those
in Figs.~\ref{residual}(d) and (e) while the component of the AB phonon mode is dominant.
The dominance of the AB phonon is seen also in the neutron strengths [Fig.~\ref{residual}(a)(c)], 
where the peak heights (red curves) are
comparable to those of the pure AB phonon mode (green curves) obtained with only the pairing interaction.
Therefore the 1.57 MeV state may be characterized as a AB-phonon-dominant mode.
Concerning the 2.06 MeV state, on the other hand,
the transition densities in Fig.~\ref{drho}(b) appear an approximate out-of-phase superposition of
those in Fig.~\ref{residual}(d) and (e) while the component of the cluster shape vibration is dominant.
Therefore it is characterized as a cluster-vibration-dominant mode.
It is not surprising that the AB phonon mode of superfluid neutron gas and the shape vibration of the cluster
couple with one another as the neutron gas and the nuclear cluster share the
same nucleon degrees of freedom. In addition the low-lying shape vibration of nuclei is known to be
influenced by the pair correlation, and hence the same is expected for the cluster shape vibration.
 A remarkable point here is that the
characters of the AB phonon mode and the cluster shape vibration remain well  in spite of the mixing.
This suggests that the AB phonon mode and the cluster shape vibration coexist and 
the coupling between the two fundamental excitation modes is rather weak
in the case of the quadrupole excitations. This is consistent with our finding in the dipole case\cite{Inakura17},
where weak coupling between the dipole AB phonon mode and the displacement motion of cluster is suggested.
  
We briefly mention other types of collective excitation modes which are identified in
the present QRPA calculation.  The third largest peak in the neutron quadrupole strength function 
[Fig.~\ref{neutron.proton.Padd.Prm} (a)] has an excitation energy of 3.03 MeV. 
The transition densities of this state, shown in Fig.~\ref{drho}(c),
suggest that this state may be interpreted as the AB phonon mode of the second harmonics with one node
at $r \approx 12$ fm with small mixing of the cluster shape vibration. A bunch of states around 
$E \approx$ 11 to 16 MeV having significant proton quadrupole strength is a high-lying
collective excitation of the cluster, which corresponds to the isoscalar giant quadrupole resonance in isolated nuclei.
A distribution of the proton strengths around 25 MeV is an analog of
the isovector giant quadrupole resonance.
Detailed analysis of these collective excitations are out of scope of the present paper, and is left for future study.

We remark here relation between the low-lying collective excitation modes found in the present calculation
and those discussed in the preceding works\cite{Khan05,Baroni2010}.
Khan et al.\cite{Khan05}  adopted the HFB plus QRPA model, which is similar to the present one. 
They discuss collective excitations in terms of the strength function for the isoscalar
quadrupole operator $Q_{IS}=Q_n + Q_p$, and found a low-lying significant peak, named
the supergiant resonance. We find in the present calculation that the isoscalar quadrupole
strength function $S(Q_{IS};E)$ looks similar to the neutron strength function $S(Q_{n};E)$
when plotted in the same scale. (This is because the proton strength is smaller by a factor of $10^{-3}$
than the neutron strength, see Figs.~\ref{drho}(a) and (b).) Therefore the supergiant resonance of
Khan et al.\cite{Khan05} may correspond to the AB-phonon-dominant mode in the present calculation.
It is likely that the cluster-vibration-dominant mode is not visible in the analysis of Khan et al. due to
their use of relatively large smoothing constant (150keV), which may hide a small peak nearby the large peak. 
Baroni et al. performed the HF plus RPA calculation, which neglects the nucleon pair correlation\cite{Baroni2010}. 
The HF plus RPA model predicts low-lying quadrupole shape vibration mode as well as the high-lying mode 
which corresponds to isoscalar 
giant quadrupole resonance of the cluster. The low-lying shape vibration mode in Ref.~\cite{Baroni2010}, however, 
exhibits some differences from the cluster-vibration-dominant mode in the present model: The former
emerges as a bump of strength distribution while the latter is a single discrete state. These differences are attributed to the
neutron pair correlation and the presence of the pair gap. Note also that due to the lack of the
pair correlation the AB phonon mode is not
expected to exist in Baroni et al. as we have seen in 
the reduced QRPA calculation including only the residual particle-hole interaction (blue curve) shown in Fig.~\ref{residual}.

\subsection{Systematics}

\begin{figure}[tbh]
\begin{center}
\includegraphics[width=0.4850\textwidth,keepaspectratio]{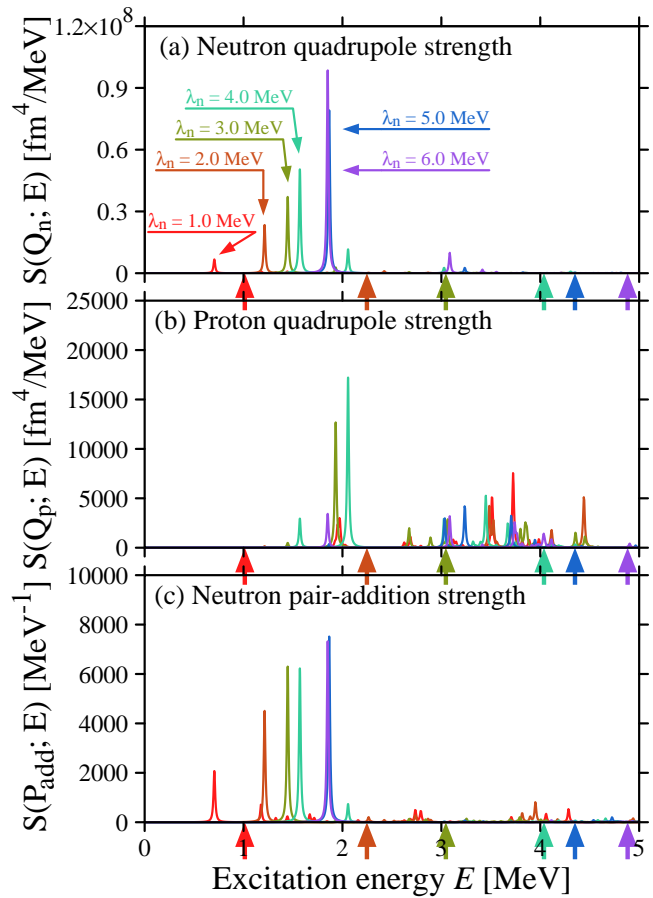}
\caption{
Strength functions (a) $S(Q_n;E)$, (b) $S(Q_p;E)$, and (c) $S(P_\mathrm{add};E)$ 
for $Z=28$ system with $\lambda_n =$ 1.0 MeV (red curves), 
2.0 MeV (brown), 3.0 MeV (yellow green), 4.0 MeV (blue green), 5.0 MeV (blue), and 6.0 MeV (purple).
Colored arrows indicate the threshold energy $2\Delta_\mathrm{ext}$ for corresponding $\lambda_n$.
See text for details.
}
\label{ABmode.quadrupole.lambda}
\end{center}
\end{figure}

We shall discuss systematic behaviour of the two kinds of low-lying collective modes by varying $\lambda_n$ and $Z$.

\begin{figure}[tbh]
\begin{center}
\includegraphics[width=0.4850\textwidth,keepaspectratio]{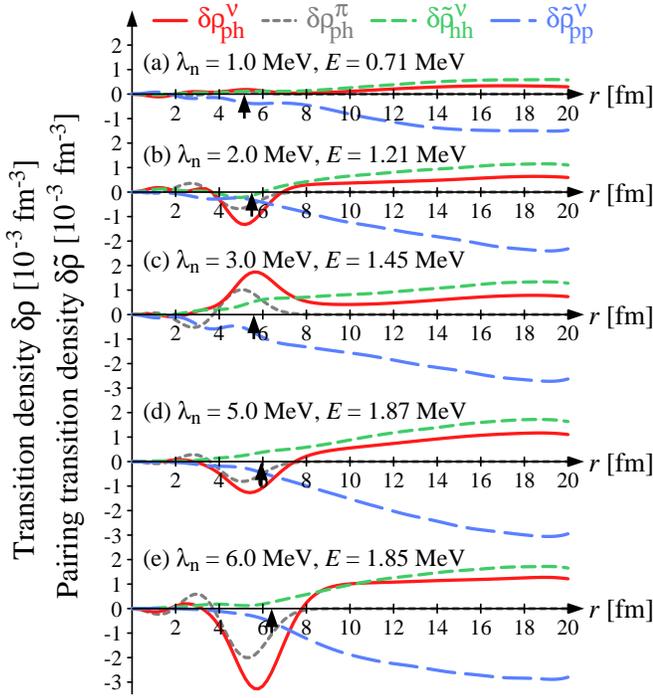}
\caption{
Transition densities 
$\delta\rho^\nu_\mathrm{ph}(r)$, $\delta\rho^\pi_\mathrm{ph}(r)$, 
$\delta\tilde{\rho}^\nu_\mathrm{pp}(r)$, and $\delta\tilde{\rho}^\nu_\mathrm{hh}(r)$ of the 
AB-phonon-dominant-mode for $Z=28$ systems with  $\lambda_n=$ 
(a) 1.0 MeV, (b) 2.0 MeV, (c) 3.0 MeV, (d) 5.0 MeV and (e) 6.0 MeV.
See text for details.
}
\label{ABdominant}
\end{center}
\end{figure}

We first examine dependence on the neutron chemical potential $\lambda_n$. 
In Fig.~\ref{ABmode.quadrupole.lambda} we plot the strength functions  
$S(Q_n;E)$, $S(Q_p;E)$, and $S(P_\mathrm{add};E)$ for $Z=28$ systems with $\lambda_n=$ 1.0, 2.0, $\cdots$, 6.0 MeV. 
It is evident in the neutron strength functions, Figs.~\ref{ABmode.quadrupole.lambda}(a)(c), that 
a low-lying collective state having large neutron strengths 
appears around 1 -- 2 MeV of excitation energy in all the cases
of $\lambda_n$.
The transition densities of these states, shown in Fig.~\ref{ABdominant}, indicate that they have the
character of the AB phonon mode in the neutron gas region $r \gesim $ 8 fm. Although admixture of 
the cluster shape vibration is seen in the transition densities,  
the proton quadrupole strength of these states is not significant (see
Fig.~\ref{ABmode.quadrupole.lambda}(b)).  Therefore all these states are interpreted as the
AB-phonon-dominant mode.  The excitation energy and the neutron strengths increase monotonically with
increase of the neutron chemical potential $\lambda_n$ .

The proton quadrupole strength, shown in Fig.~\ref{ABmode.quadrupole.lambda}(b),
exhibits different behaviours. The low-lying collective state with enhanced proton strength are found
only for $\lambda_n=$ 3.0 MeV and 4.0 MeV. 
These are the cluster-vibration-dominant mode discussed in the subsection just above.
There exists no cluster-vibration-dominant mode in the low-lying region 
in the cases of $\lambda_n=$ 2.0 MeV, 5.0 MeV and 6.0 MeV.
(In case of  $\lambda_n=$ 1.0 MeV, we find a state which has partial character of the cluster vibration
at $E=$ 1.18 MeV, which lies above the continuum threshold $2 \Delta_\mathrm{ext}$.)

\begin{figure}[tb]
\begin{center}
\includegraphics[width=0.49990\textwidth,keepaspectratio]{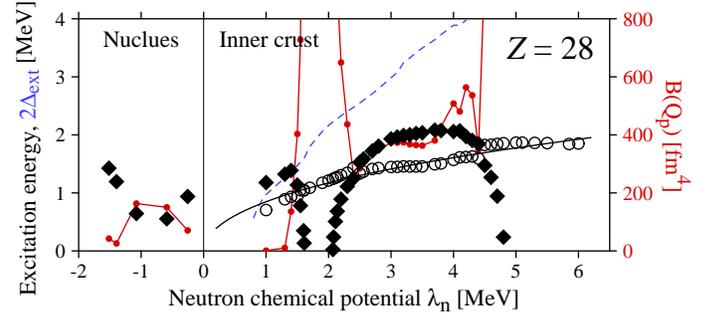}
\caption{Excitation energy $E_{AB}$ of the lowest AB-phonon-dominant mode (open circle) and
the energy $E_{cl}$ of the cluster-vibration-dominant mode (filled diamond) for $Z=$ 28 systems,
plotted as a function of the neutron chemical potential $\lambda_n$. 
Blue dashed curve is the threshold energy $2\Delta_\mathrm{ext}$.
Red dots connected with line denote the proton quadrupole strengths $B(Q_p)$ of the cluster-vibration-dominant mode.
The results for near-drip-line nuclei $^{80,82,\cdots,88}$Ni are also plotted.
The thin black line is the hydrodynamic estimate $\hbar\omega_{\rm hyd}$ of the quadrupole
AB phonon mode in uniform superfluid neutron gas in the same Wigner-Seitz cell. 
}
\label{ABmode.Z28}
\end{center}
\end{figure}

Figure~\ref{ABmode.Z28} shows more detailed information on the excitation energy $E_\mathrm{AB}$ of 
the AB-phonon-dominant mode 
and the energy $E_\mathrm{cl}$ of 
the cluster-vibration-dominant mode, plotted as a function of $\lambda_n$. 
The proton quadrupole strength $B(Q_p)$ of the cluster-vibration-dominant mode is also plotted.
For comparison we also plot the excitation energy and $B(Q_p)$ of the $2_1^+$ state of the near-drip-line nuclei $^{80-88}$Ni 
which have negative $\lambda_n$.

The excitation energy $E_\mathrm{AB}$ of the AB-phonon-dominant mode has smooth dependence on $\lambda_n$.
It agrees well with an  hydrodynamic estimate of the excitation energy of the quadrupole AB mode, $\hbar\omega_\mathrm{hyd}$,
which is obtained assuming  that the phonon amplitude is expressed by the spherical Bessel function, $\phi(r) \propto j_2(qr)$
with boundary condition $d\phi(r)/dr =0$ at $r=R_\mathrm{WS}=20$ fm.\footnote{
In this estimate, we assume $\phi(r) \propto j_2(qr)$ together with the boundary condition $\left. d\phi(r)/dr \right|_{r=R_\mathrm{box}}=0$.
Combination of the phonon dispersion relation $\omega = c_\mathrm{ph} q$ and 
the hydrodynamic estimate of the phonon velocity $c_\mathrm{ph}=\sqrt{(\partial\lambda_n/\partial\rho)\rho/m}$ 
gives an estimated excitation energy $\hbar\omega_\mathrm{hyd}$.
Here, we use $\rho(\lambda_n)$ evaluated for uniform neutron matter with the same Skyrme functional SLy4.
}
This indicates that the presence of the nuclear cluster 
affects only to a small extent the excitation
energy of the AB-phonon-dominant mode. It is also consistent with our observation  
in Figs.~\ref{neutron.proton.Padd.Prm} and \ref{residual} that the coupling between the quadrupole
shape vibration of cluster and the quadrupole AB phonon is weak.
For $0< \lambda_n <$ 1 MeV, we do not plot the results since 
the collective modes do not appear at low excitation energies 
below $2\Delta_\mathrm{ext}$.
Note however that the AB phonon mode would appear below 
$2\Delta_\mathrm{ext}$ if we adopted a large Wigner-Seitz cell.

The cluster-vibration-dominant mode displays strong, non-monotonic and oscillatory $\lambda_n$-dependence.
As $\lambda_n$ increases slightly above 1.3 MeV the excitation energy $E_\mathrm{cl}$ 
drops off sharply. The $B(Q_p)$ value increases dramatically. This behaviour reminds us
 of the softening of low-lying quadrupole 
vibration in isolated nuclei, and the same appears to happen for the nuclear cluster in the 
Wigner-Seitz cell. 
The cluster-vibration-dominant mode is not found
in the interval of $\lambda_n \approx 1.6 - 2.0$ MeV, and it suggests instability of
the spherical nuclear cluster  
 toward the quadrupole deformation. (In this case 
 the QRPA energy of the quadrupole shape vibration is expected to be  ``imaginary".)  
In the interval of $\lambda_n \approx  2.1 \sim 4.8$ MeV, the cluster-vibration-dominant mode emerges at finite 
excitation energy.
The softening is seen above $\lambda_n \gesim 4.5$ MeV and the instability occurs again in the interval
$\lambda_n \approx 4.8 \sim 6$ MeV.

\begin{figure}[tb]
\begin{center}
\includegraphics[width=0.4380\textwidth,keepaspectratio]{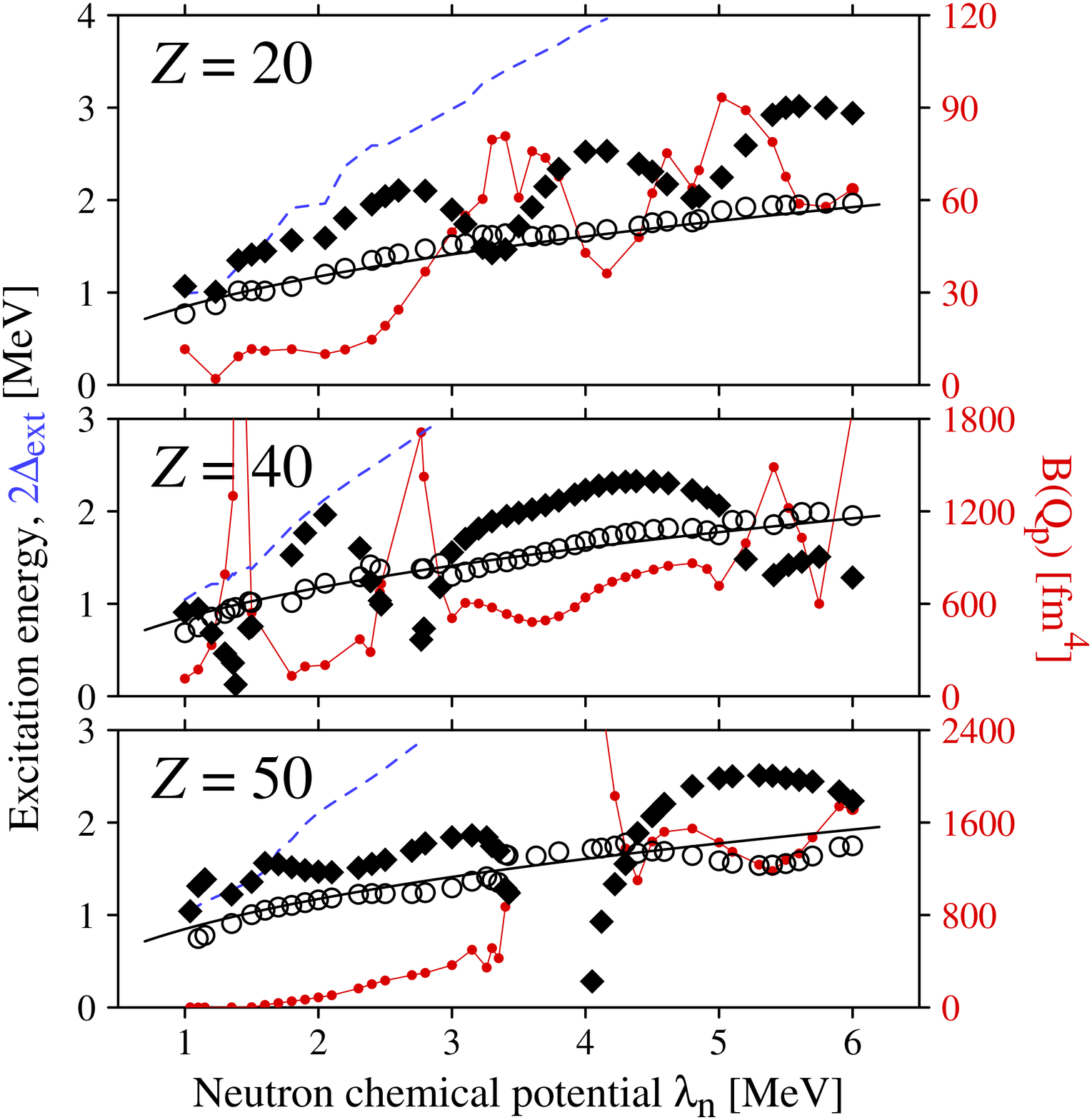}
\caption{The same as Fig.~\ref{ABmode.Z28}, but for systems with $Z=$ 20, 40, and 50.
}
\label{ABmode.Zall}
\end{center}
\end{figure}

Figure~\ref{ABmode.Zall} shows the same plot as Fig.~\ref{ABmode.Z28}, but for systems
with $Z=$ 20, 40 and 50. The energy of the AB-phonon-dominant mode agrees well with the
hydrodynamical estimate while the strong and oscillatory $\lambda_n$-dependence of the
cluster-vibration-dominant mode is present in all the cases.

\subsection{Resonance shell effect on cluster shape vibration}

\begin{figure}[tb]
\begin{center}
\includegraphics[width=0.480\textwidth,keepaspectratio]{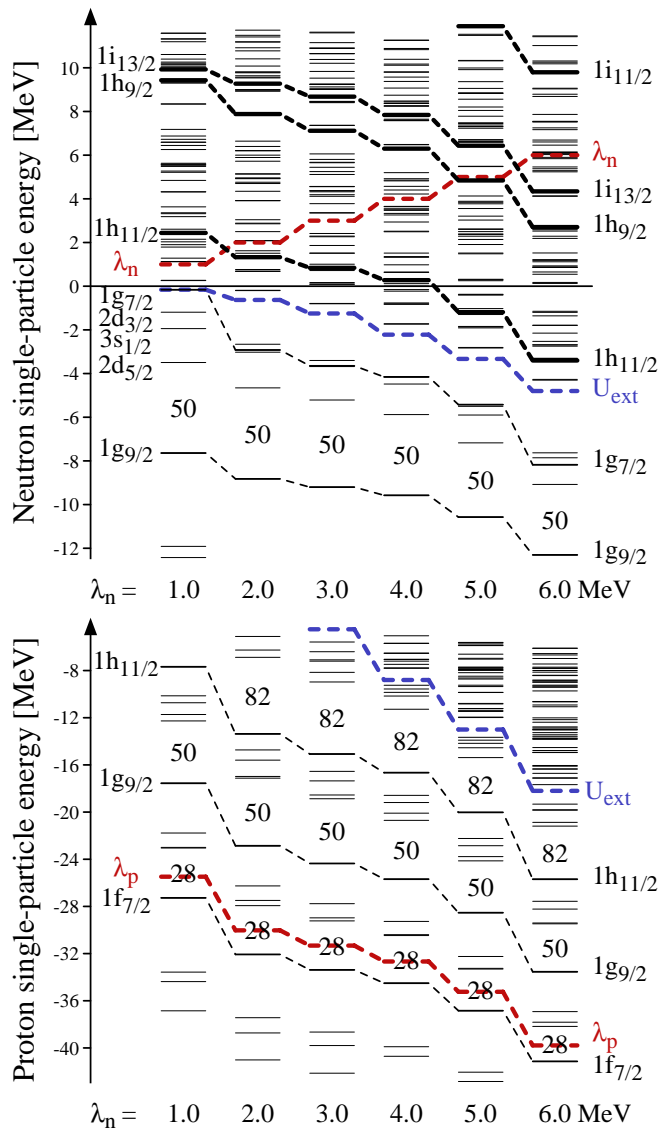}
\caption{(a) Neutron  single-particle energies for $Z=$ 28 systems with $\lambda_n=$ 1.0, 2.0, $\cdots$, 6.0 MeV.
Thick back bars connected with dashed lines denote the resonant single-particle states with quantum
numbers  $1h_{11/2}$, $1h_{9/2}$,  $1i_{13/2}$, and $1i_{11/2}$. Bound high-$j$ orbits are also connected with
thin dashed lines for reference.
Red horizontal bars connected with dashed line denote the neutron Fermi energy (chemical potential) $\lambda_n$, and blue 
bars the neutron Hartree-Fock potential  $U_\mathrm{ext}$ in the region of neutron gas.
 (b) The same as (a), but for protons.
Numbers 28, 50 and 82 inserted in levels are the spherical magic numbers. 
}
\label{spe}
\end{center}
\end{figure}

The strong and oscillatory $\lambda_n$-dependence suggests possible shell effect in the cluster-vibration-dominant mode.
However, the conventional idea of the shells characterized by the magic numbers $N=$ 50, 82, 126, etc. is not
appropriate here as  neutron single-particle orbits around the
neutron Fermi energy $\lambda_n$ are unbound states having a quasi-continuum spectrum.

In order to clarify the suggested shell effect, we look into the single-particle eigenstates in the Hartree-Fock potential.
Figure~\ref{spe} shows the neutron and proton single-particle energies for the $Z=$ 28 systems with
$\lambda_n= 1.0, 2.0, \cdots, 6.0$ MeV. The thick blue bar indicates the depth $U_{\rm ext}$ of the Hartree-Fock
potential in the region of neutron gas, namely in the external region outside the nuclear cluster. The single-particle
states with energy below $U_{\rm ext}$ are localized states bound to the nuclear cluster while those above
$U_{\rm ext}$ are unbound states, which also have discrete energies because of the Wigner-Seitz approximation with
finite cell radius. 
Concerning the bound single-particle states, the single-particle energy gaps at 
the neutron number $N=50$ and at the proton 
numbers $Z=$ 28, 50 and 82 are clearly visible.   For unbound single-particle states, on the other hand,
the single-particle energy spacings are much smaller than those of bound states, and the spacings approach zero 
if we take larger Wigner-Seitz radius. 
We have examined wave functions of all the unbound 
single-particle states of neutrons. Among the discretized unbound states, we find a few characteristic states
whose wave function has large amplitude inside the nuclear cluster ($r \lesim 6$ fm), at least larger than amplitude
in the exterior region.
Such states correspond to resonant single-particle states and they are marked with thick lines in Fig.~\ref{spe}(a). 
The resonant single-particle states are high-$l$ orbits with no radial node inside the cluster, 
which can be labeled as $1h_{11/2}$, $1h_{9/2}$,
$1i_{13/2}$ etc. 
Note that these high-$l$ orbits have  large
centrifugal barriers which allow
resonances. The
barrier height for $l=5$ partial waves ($h_{11/2,9/2}$) is about 10 MeV in the case of $\lambda_n=5.0$ MeV.

We found that these resonant high-$l$ orbits are the origin 
of the shell effect seen in the cluster shape vibration. 

Thick red bars in Fig.~\ref{spe}(a) indicate the neutron Fermi energy (chemical potential) $\lambda_n$. 
As $\lambda_n$ increases, the relative position of $\lambda_n$ moves upward
 with respect to the spectrum of unbound
neutron single-particle states. Then, the neutron Fermi energy $\lambda_n$ crosses
with the resonant $1h_{11/2}$ neutron state 
 at $\lambda_n \approx 1.8$ MeV while the other resonant 
$1h_{9/2}$ and $1i_{13/2}$ orbits crosses with $\lambda_n$ 
at $\lambda_n \approx 5.0$ and 5.5 MeV, respectively.
It is seen that the softening and the instability of the cluster shape vibration mode occurs 
when these crossings occur.

With these observations, we infer the mechanism of the shell effect seen in the
quadrupole shape vibration of cluster. A resonant high-$l$ orbit has wave function
concentrated  in the region of the nuclear cluster, and hence it
may play a similar role of a single-$j$ shell in nuclear structure problems. 
Furthermore we can assume that non-resonant unbound single-particle orbits do not play 
major roles as they are single-particle states contributing mainly to neutron gas, but not to
the nuclear cluster.
In this situation  enhancement of the quadrupole collectivity occurs when  the
Fermi energy is located close to the energy of the single-$j$ shell (the
resonant high-$l$ state in the present case)  and the shell
is approximately half-filled.  This shell effect  caused by the resonant single-particle states 
is very different from the conventional 
one, in which 
the shells refer to bound orbits in nuclei and the associated magic numbers
20, 28, 50, 82, etc. It also differs from the
continuum shell effect\cite{BulgacMagierski2001,MagierskiHeenen2002} which is
present in the level density of  the non-resonant unbound states. 
 To distinguish it from the existing concepts of the shell effect,  we
shall call it {\it resonance shell effect}. Note that Baroni et al. \cite{Baroni2010} points out
importance of neutron resonant high-$l$ single-particle states as one of 
major contributors to the low-lying and high-lying quadrupole modes. Their discussion
is related to the present resonance shell effect,  but the roles of the Fermi energy
and the pair correlation need to be noticed to explain
the oscillatory  behaviour of  the quadrupole 
collectivity as a function of the neutron Fermi energy. 

The strong and oscillatory $\lambda_n$-dependence of the
cluster-vibration-dominant mode in the $Z=$ 20, 40 and 50 systems, shown in
Fig.~\ref{ABmode.Zall}, is also nicely explained with the resonance shell effect.
The softening of the shape vibration around $\lambda_n \approx$ 3.3 MeV and 4.8 MeV 
in the $Z=20$ system can be related to $1g_{7/2}$ 
and $1h_{11/2}$ resonant states of neutrons, respectively.
For the softening around $\lambda_n \approx$ 1.4 MeV, 2.6 MeV  and 5.4 MeV 
in the $Z=40$ system,  neutron $1h_{9/2}$, $1i_{13/2}$, and $1i_{11/2}$ resonant states, respectively,
are the keys. The instability around $\lambda_n \approx$ 3.7 MeV in 
 the $Z=50$ system is due to neutron $1i_{11/2}$  resonant state.

In the present study, we have analyzed the systems with $Z=$ 20, 28, 40 and 50
in which the protons have closed-shell configurations. In realistic calculations of
inner crust\cite{Negele73,Baldo2005,Baldo2006,Grill11}, the predicted proton number of 
nuclear cluster is not limited to these
magic numbers, but rather the protons often have open shell configurations
where the proton pair correlation  also takes part in. In such cases, the
quadrupole collectivity of the cluster shape vibration is even more enhanced
due to the contribution of open shell protons
than in the present case of proton closed shell, and there will be more possibilities
of the quadrupole deformation of nuclear clusters in situations where
the resonance shell effect is expected. It is also remarked that 
the occurrence of quadrupole-deformed
nuclear cluster caused by the resonance shell effect is expected even 
at low densities (as well as high densities) of superfluid neutron gas.
This is in contrast to the deformation mechanism originating from the
Coulomb lattice energy, which is responsible for
the manifestation of various pasta structures\cite{Pethick-Ravenhall95,
RavenhallPethick1983, HashimotoSeki1984, Oyamatsu1993}. It is interesting to
explore deformed clusters at low neutron-gas densities, $0 < \rho_{n,{\rm ext}} \lesim \rho_0/10$,
by means of
quantum mechanical approach such as the HFB and the 
Bogoliubov-de-Genne-Kohn-Sham density functional models. Note that the previous HF calculations
analyzing  deformed clusters\cite{MagierskiHeenen2002,GegeleinMuther2007,NewtonStone2009, PaisStone2012, FattoyevHorowitz2017}
focus on the pasta structures in the bottom layers of inner crust where the neutron-gas density or
the average baryon density is rather high $\rho \gesim \rho_0/10$.

\section{Conclusions}

We have investigated collective excitations in the inner crust
of neutron stars in the framework of the nuclear density functional theory.
We consider configurations where spherical clusters are immersed in superfluid neutron gas, and
describe low-lying quadrupole collective excitations around the cluster by means of
the HFB plus QRPA model formulated in a spherical Wigner-Seitz cell.
We adopt the Skyrme SLy4 functional and the density-dependent delta pairing interaction.
Numerical analysis is performed by varying the neutron chemical potential (Fermi energy)
$\lambda_n =1 - 6$ MeV changing the density of superfluid neutrons and
the proton number of the cluster.

Our calculation demonstrates that there coexist the Anderson-Bogoliubov (AB) phonon mode of neutron superfluid and the
quadrupole shape vibration of the cluster, and these two fundamental modes mix with each other only weakly.
The AB-phonon-dominant-mode, i.e.  the AB phonon mode with small mixing of the
quadrupole shape vibration of cluster, emerges systematically, and its excitation energy exhibits 
smooth and monotonic dependence on $\lambda_n$, which agrees fairly well with the hydrodynamical estimate
of the AB phonon in uniform neutron superfluid. 
The results point to the weak coupling
between the quadrupole AB phonon mode of neutron gas and the quadrupole shape vibration of clusters.
This is  consistent with our previous study\cite{Inakura17} suggesting that 
the dipole AB phonon and the displacement motion of clusters is weak.

Another remarkable finding is that the excitation energy of the cluster shape vibration varies in the range of 
$0-3$ MeV with strong and oscillatory $\lambda_n$-dependence, in contrast to the AB phonon mode. 
This behavior is traced to a new kind
of shell effect, named the resonance shell effect, which is associated with 
unbound but resonant single-particle states of neutrons orbiting around the cluster with large orbital angular momentum.
The resonance shell effect causes enhancement of collectivity, softening and  even instability of the quadrupole shape  degrees
of freedom of clusters. This suggests that occurrence of quadrupole-deformed nuclear cluster is possible in any layers of inner crust
if the resonance shell effect induces the instability.

\section*{Acknowledgments}

This work was supported by JSPS KAKENHI Grant Number JP17K05436, and by Grant-in-Aid for Scientific 
Research on Innovative Areas, No. 24105008, by The Ministry of Education, 
Culture, Sports, Science and Technology, Japan. \\

\end{document}